\documentclass[12pt]{article}
\usepackage{a4wide}
\usepackage{amsmath}
\usepackage{axodraw}
\usepackage{graphicx}
\newcommand{\meno}{\textrm{--}\,}

\newcommand{\lqcd}{\Lambda_{\textrm{\small{QCD}}}}
\def\gev{\,\mathrm{Ge\kern-0.1em V}}

\begin{document}
\begin{flushright}
SHEP 01/01\\ 
\end{flushright}
\begin{center}
{\Large\textbf{Lattice \boldmath{$B$}-Physics}}~\footnote{Invited lecture
presented at the 7th International Conference on $B$-Physics at Hadron
Machines, \textit{Beauty 2000}, Magaan, Sea of Galillee, Israel, 13-18
September, 2000.}\\ 

\vspace{0.3in}
\textbf{C T Sachrajda}\\ 
Department of Physics and Astronomy\\ 
University of Southampton\\
Southampton SO15 7PW\\
UK
\end{center}

\begin{abstract}
I review the status of lattice simulations relevant for phenomenological
studies of $B$-physics. Results for much-studied quantities such as
$f_B$, $B_B$ and form-factors for semileptonic decays are presented as
well as those for quantities which have begun to be studied only recently
(such as $B$-lifetimes). The improvement in the precision of the
determinations of the mass of the $b$-quark, which has been made possible
by new results for perturbative coefficients, is discussed. Finally
I describe new ideas being developed with the aim of making computations
of two-body non-leptonic decays possible.
\end{abstract}

\section{Introduction}
\label{sec:intro}
At this conference we have seen a huge amount of experimental data
being presented on $B$-physics, data from which we would like to
extract fundamental information about the parameters of the standard
model and CP-violation, and (by overconstraining the comparison with
the standard model) to search for signals of new physics. Of course
much more data will become available in the coming years. A major
difficulty in achieving the above goals is our inability to quantify
non-perturbative strong interaction effects. Lattice QCD provides the
opportunity for evaluating these effects \textit{ab initio}, and there
is indeed a wide-ranging program of numerical simulations of
$B$-physics. In this lecture I will briefly review some of the recent
progress in lattice studies of $B$-physics.

It is not appropriate for me in this talk to discuss technical aspects
of lattice calculations in any detail, I will therefore concentrate on
presenting results and discussing the potential for future
calculations. I start with some basic facts about lattice calculations
in $B$-physics. In any simulation the computing resources are limited,
and one therefore has to compromise when minimising
competing systematic errors, those due to the granularity of the
lattice (discretization effects) and finite-volume errors. In current
simulations, this typically leads to a choice for the lattice spacing
($a$) in the range
given by $a^{-1}\sim 2$\,--\,4\,GeV, i.e. $a$ is larger than the Compton
wavelength of the $b$-quark. This means that we cannot study the
propagation of a physical $b$-quark directly and either have to
i) use effective theories, such as the Heavy Quark Effective
Theory (HQET) or Non-Relativistic QCD (NRQCD) or
ii) calculate physical quantities with the heavy-quark mass
$m_Q$ in the region of $m_c$ (the mass of the charm-quark), and perform
the extrapolation to $m_Q=m_b$.

The precision of lattice calculations is limited by systematic
uncertainties. These can be overcome in principle with increased
computing resources and much effort is being devoted to reducing these
uncertainties (and in particular to eliminating \textit{quenching},
i.e.  to include vacuum polarisation effects). Most, but not all, of
the results presented below have been obtained in the quenched
approximation. It should be remembered however, that even in
unquenched simulations the masses of the sea quarks are
large~\footnote{For example we have yet to simulate with sufficiently
  light quarks so that the $\rho\to\pi\pi$ decay is kinematically
  possible.}. Controlling the effects of sea-quarks is perhaps the
major challenge for the lattice community and in recent years we have
began to address this challenge.

From lattice simulations we know how to calculate matrix elements of
the form $\langle H\,|\,O(0)\,|\,0\rangle$ and
$\langle\,H_1\,|\,O(0)\,|\,H_2\,\rangle$, where $H$ and $H_{1,2}$ are
hadrons and $O(0)$ is some local composite operator (several examples
of such matrix elements will be discussed below). We are not yet in a
position to compute matrix elements in which there are two or more
hadrons in the initial or final state, and are therefore unable to
compute $B\to H_1H_2$ decays amplitudes, but progress is
being made in this area (see sect.~\ref{sec:nonlept} below).

The plan of this lecture is as follows. In the next section I briefly
review the results for some \textit{standard} physical quantities:
leptonic decay constants $f_B$ and $f_{B_s}$; the $B_B$-parameters of
$\bar B$-$B$ mixing and form-factors for semileptonic decays.
Sect.~\ref{sec:inclusive} contains a discussion of computations of
inclusive decays and lifetimes. Results have been recently obtained
for these quantities but they have not been studied as extensively as
those in sect.~\ref{sec:fb}. New calculations of perturbative
coefficients have enabled the mass of the $b$-quark to be determined
from lattice simulations with considerably better precision and this
is reviewed in sect.~\ref{sec:mb}. I then describe recent attempts to
develop techniques to enable two-body non-leptonic decays to be
studied on the lattice (sect.~\ref{sec:nonlept}). Finally I summarise
the main conclusions in sect.~\ref{sec:concs}.

\section{Decay Constants, B--Parameters and Form-Factors}
\label{sec:fb}

Leptonic decay constants $f_B$ and $f_{B_s}$, the $B$--parameters of
$B$--$\bar B$ mixing and the form-factors for semileptonic $B$-decays
have been computed in lattice simulations for over ten years now. In this
section I briefly discuss each of these quantities in turn.

\subsection{$\mathbf{f_B}$}
\label{subsec:fb}

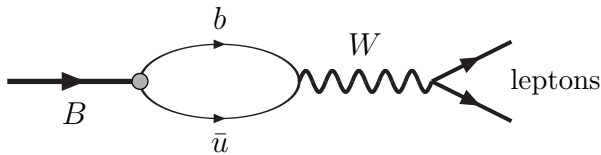
\begin{figure}[t]
\begin{center}
\begin{picture}(300,40)(-120,-20)
\Text(-55,-8)[t]{$B$}
\Text(0,20)[b]{$b$}\Text(0,-20)[t]{$\bar u$}
\Text(55,10)[b]{$W$}
\Text(110,0)[l]{\small{leptons}}
\ArrowLine(-0.5,14)(0.5,14)\ArrowLine(-0.5,-14)(0.5,-14)
\SetWidth{2}\ArrowLine(-80,0)(-30,0)\SetWidth{0.5}
\Oval(0,0)(14,30)(0)\GCirc(-30,0){3}{0.7}
\SetWidth{1.5}
\Photon(30,0)(80,0){4}{5}\ArrowLine(80,0)(110,15)
\ArrowLine(80,0)(110,-15)
\end{picture}
\caption{Diagrammatic representation of a leptonic decay of a
$B$--meson.\label{fig:fb}} 
\end{center}
\end{figure}

The strong interaction effects in leptonic decays of $B$--mesons (see
fig.\ref{fig:fb}) are contained in the matrix element 
\begin{equation} 
\langle\,0\,|\,A_\mu(0)\,|\,B(p)\,\rangle=i f_B p_\mu\ .
\label{eq:fbdef}\end{equation}
Lorentz and Parity Invariance imply that all the non-perturbative QCD
effects are para\-me\-trized in terms of a single number, $f_B$, defined
in eq.(\ref{eq:fbdef}).

Claude Bernard, the reviewer at the Lattice 2000 symposium~\cite{cb}
summarised the current status of the results as~\footnote{Ref.~\cite{cb}
contains a detailed critical analysis of all the lattice results
for the quantities considered in this section.}: 
\begin{center}
\vspace{-0.3in}
\begin{tabular}{lll}\\ 
&Full Theory&$N_f=0$\\ \hline
$f_B$ & $200\pm 30$\,MeV & $175\pm 20$\,MeV\\ 
$f_{B_s}/f_B$&$1.16\pm 0.04$&$1.15\pm0.04$
\end{tabular}
\end{center}
At last year's conference the reviewer S.Hashimoto stated that
\textit{all available data is consistent with the following
estimates:}~\cite{hashimoto}
\begin{center}
\vspace{-0.3in}
\begin{tabular}{lll}\\ 
&$N_f=2$&$N_f=0$\\ \hline
$f_B$ & $210\pm 30$\,MeV & $170\pm 20$\,MeV\\ 
$f_{B_s}$ & $245\pm 30$\,MeV & $195\pm 20$\,MeV\\ 
$f_{B_s}/f_B$&$1.16\pm 0.04$&$1.15\pm0.04$
\end{tabular}
\end{center}

These results have been largely stable for some time. The last time
that I performed a compilation was in 1997 together with Jonathan
Flynn~\cite{fs} and reported (on what were mainly quenched calculations):
\begin{equation}
f_B = 170\pm 35\,\textrm{MeV},\ \ 
f_{B_s} = 195\pm 35\,\textrm{MeV},\ \ 
\frac{f_{B_s}}{f_B}=1.14 \pm 0.08.
\label{eq:fs}\end{equation}
Although the results have been stable, the errors have been decreasing
very slowly (if at all). The errors will not decrease significantly
until we begin to get a serious control of quenching errors. In
unquenched calculations, the value of the lattice spacing typically
varies by 10\% or so depending on the physical quantity which is used to
set the scale. It is therefore not really possible to determine
dimensionful quantities such as $f_B$ with a better precision  than about
10\%.

In the last two years we have began to study quenching effects,
although it must be remembered that at present the masses of the
sea-quarks are still relatively heavy (typically a little smaller that
the strange quark mass). These effects can only be studied
meaningfully if all other variables are kept constant. Using results
from the MILC~\cite{milc}, CPPACS~\cite{cppacs} and NRQCD~\cite{nrqcd}
collaborations, C.~Bernard estimates that there is about a 20~MeV
increase in the value of $f_B$ in going from the quenched
approximation to the $N_f=2$ case (using $f_\pi$ to set the
scale)~\cite{cb}. The value of $f_{B_s}/f_B$ appears to be the same in
the two cases.

Although the errors have not been reduced substantially, there have
been many systematic checks on the stability of the results. For
example discretization errors have been studied by using improved
actions and/or extrapolating the results to the continuum limit. It
had been suggested that $f_B$ may be significantly lower than the
results presented above because of discretization errors (this
suggestion was based on some extrapolations to the chiral limit), but
this has not survived more careful analyses~\cite{cb}.

When lattice computations of the leptonic decay constants of heavy mesons were
beginning there were no experimental data. In 1997 we compiled the (quenched)
lattice predictions for the decays constants of charmed mesons as~\cite{fs}: 
\begin{equation} f_D= 200\pm 30\,\textrm{MeV\ \ and\ \ } 
f_{D_s}= 220\pm 30\,\textrm{MeV}.\label{eq:fdlattice}\end{equation}
The Particle Data Group's review this year quotes $f_{D_s}=280\pm
19\pm 28\pm34$\,MeV~\cite{chada} as the best experimental result (and
recent results, still with large errors, are compatible with this). It
will be interesting to observe future lattice and experimental results as
the errors in both decrease.

Finally let me mention that lattice computations have shown that there
are large corrections to the HQET scaling law for the decay constants of
heavy-light pseudoscalar mesons~$P$ 
\begin{equation}
f_P\sim \frac{1}{\sqrt{m_P}}.
\label{eq:fbscaling}\end{equation}

\subsection{$\mathbf{B_B}$:}
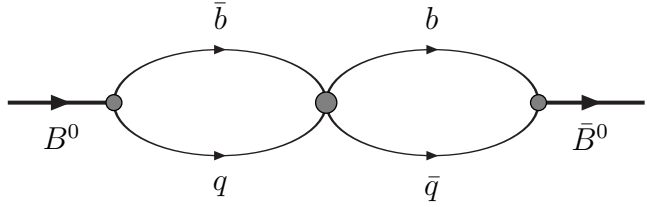
\begin{figure}[t]
\begin{center}
\begin{picture}(300,80)(-150,-30)
\Oval(-40,0)(20,40)(0)\Oval(40,0)(20,40)(0)
\SetWidth{1.5}\ArrowLine(-120,0)(-80,0)
\ArrowLine(80,0)(120,0)
\SetWidth{0.5}
\ArrowLine(-40.5,20)(-39.5,20)\ArrowLine(-40.5,-20)(-39.5,-20)
\ArrowLine(39.5,20)(40.5,20)\ArrowLine(39.5,-20)(40.5,-20)
\GCirc(-80,0){3}{0.5}\GCirc(80,0){3}{0.5}
\GCirc(0,0){4}{0.5}
\Text(-100,-8)[t]{$B^0$}\Text(100,-8)[t]{$\bar B^0$}
\Text(-40,28)[b]{$\bar b$}\Text(40,28)[b]{$b$}
\Text(-40,-28)[t]{$q$}\Text(40,-28)[t]{$\bar q$}
\end{picture}
\end{center}
\caption{Diagramatic representation of $B^0$-$\bar B^0$ mixing.
\label{fig:mixing}}
\end{figure}

The relevant matrix element for $B^0$-$\bar B^0$ mixing is:
\[M(\mu)\equiv\langle\,\bar B^0\,|\,\bar b\gamma_\mu(1-\gamma_5)q\ 
\bar b\gamma^\mu(1-\gamma_5)q\,|\,B^0\,\rangle\,,\]
where $q$ represents $d$ or $s$. This is illustrated in 
fig.~\ref{fig:mixing}.

Following the conventions introduced in kaon physics, 
B-parameters are defined by
\begin{equation}
 M(\mu)=\frac{8}{3}f_B^2m_B^2B_B(\mu).
\label{eq:bbdef}\end{equation}
$B_B(\mu)$ is scheme and scale-dependent, so it is convenient to
define a renormalization scheme and scale- independ\-ent (up to NLO)
quantity
\begin{equation}
\hat B_B^{\textrm{nlo}}=\alpha_s(\mu)^{2/\beta_0}\left[
1\,+\,\frac{\alpha_s(\mu)}{4\pi}\,J_{n_f}\right]\ B_B(\mu).
\label{eq:bbhatdef}\end{equation}
where $J_{n_f}$ is a known constant. Many groups have studied mixing
in lattice simulations and C.\,Bernard, the reviewer at this year's
lattice conference, summarised the results as:
\begin{eqnarray}
\hat B_{B_d} &=& 1.30\pm 0.12\pm 0.13,\hspace{0.5in} 
\hat B_{B_s}/\hat B_{B_d}=1.00\pm 0.04\,,\\ 
f_B\sqrt{\hat B_{B_d}} &=&230 \pm 40\,\textrm{MeV},\hspace{0.5in} 
\xi\equiv \frac{f_{B_s}\sqrt{\hat B_{B_s}}}
{f_B\sqrt{\hat B_{B_d}}}=1.16\pm 0.05\,.
\end{eqnarray}
The second error for $B_{B_d}$ is an estimate of the error due to
quenching. $\xi$ is a particularly important quantity for studies of
the unitarity triangle. It should be noted that the error on $\xi$ is
not a small one from the lattice perspective, indeed given its central
importance in phenomenological studies most groups tend to be cautious
in determining this error. The reason that it should not be
considered small is that the quantity which is being computed is
$\xi-1=0.16 \pm 0.05$, i.e. it has a quoted error of over 30\%. In my
judgement therefore the uncertainty in the value of $\xi$ should not be 
further inflated in phenomenological studies. The results for $\xi$ are
stable, and there is no evidence that they change as sea-quarks
are introduced.

For comparison with the results quoted above I present also those
quoted in our 1997 review~\cite{fs}:
\begin{eqnarray}
\hat B_{B_d}&=&1.4\pm 0.1, \hspace{0.5in} 
\hat B_{B_d}/\hat B_{B_s}=1.01\pm 0.01\,,\\ 
f_B\,\sqrt{\hat B_{B_d}}&=&201(42)\,\textrm{MeV}\,,\hspace{0.5in}   
\xi\equiv\frac{f_{B_s}\sqrt{\hat B_{B_s}}}
{f_{B}\sqrt{\hat B_{B}}}= 1.14\pm 0.08.
\end{eqnarray}

These results have been obtained with propagating heavy quarks
(i.e. by extrapolating results obtained with $m_Q\simeq m_c$).  In the
static approximation the perturbative corrections are very large and
hence these have not given us much useful information to date.

\subsection{Exclusive Semi-Leptonic \boldmath{$B$}-Decays}
\label{subsec:semil}

\begin{figure}[h]
\begin{center}
\begin{picture}(380,100)(-130,-25)
\Oval(0,0)(24,60)(0)
\SetWidth{1.5}\ArrowLine(-120,0)(-60,0)\ArrowLine(60,0)(120,0)
\Photon(0,24)(50,54){4}{5}
\ArrowLine(50,54)(80,74)\ArrowLine(50,54)(80,34)
\SetWidth{0.5}
\GCirc(-60,0){3}{0.5}\GCirc(60,0){3}{0.5}
\Text(-90,-8)[t]{$B$}\Text(90,-8)[t]{\small{$D^{(\ast)},\pi,\rho$}}
\Text(75,54)[l]{\small{leptons}}
\Text(-30,26)[b]{\small{$b$}}
\Text(35,22)[bl]{\small{$c,u$}}
\Text(135,0)[l]{$\Rightarrow\ \ V_{cb}\,,\, V_{ub}$}
\end{picture}
\caption{Diagramatic Representation of Semileptonic B-Decays.
\label{fig:semil}}
\end{center}
\end{figure}
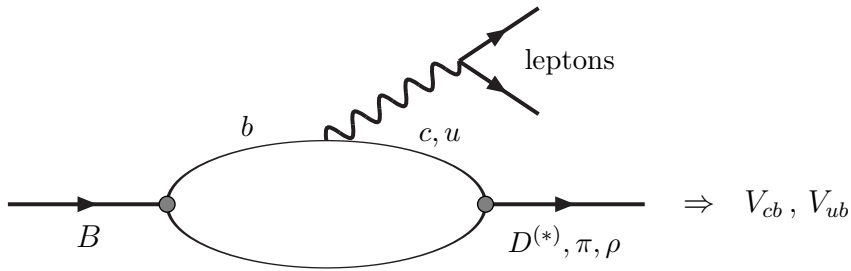

Exclusive semileptonic decays of $B$-mesons are being used to determine
the $V_{cb}$ and $V_{ub}$ CKM-matrix elements. Diagramatically the
amplitudes can be represented by the diagram in fig.~\ref{fig:semil}.
Lorentz and parity invariance imply that the amplitudes can be
expressed in terms of invariant form factors. For example for the decays
into a pseudoscalar $P$ ($B\to\pi$ or $B\to D$ decays),
in the helicity basis, 
\begin{equation}
\langle\,P(p_P)\,|V_\mu(0)\,|\,B(p_B)\,\rangle=
f^0(q^2)\,\frac{M_B^2-M_P^2}{q^2}q_\mu+ 
f^+(q^2)\,\left\{(p_B+p_P)_\mu-\frac{M_B^2-M_P^2}{q^2}q_\mu
\right\}.
\end{equation}
As a result of parity invariance only the vector component (from the
$V-A$ weak current) contributes in this case. When the final-state
hadron is a vector meson ($\rho$ or $D^\ast$) the amplitude is written
in terms of four invariant form factors.

\begin{figure}[t]
\hbox to\hsize{%
\includegraphics[width=0.42\hsize]{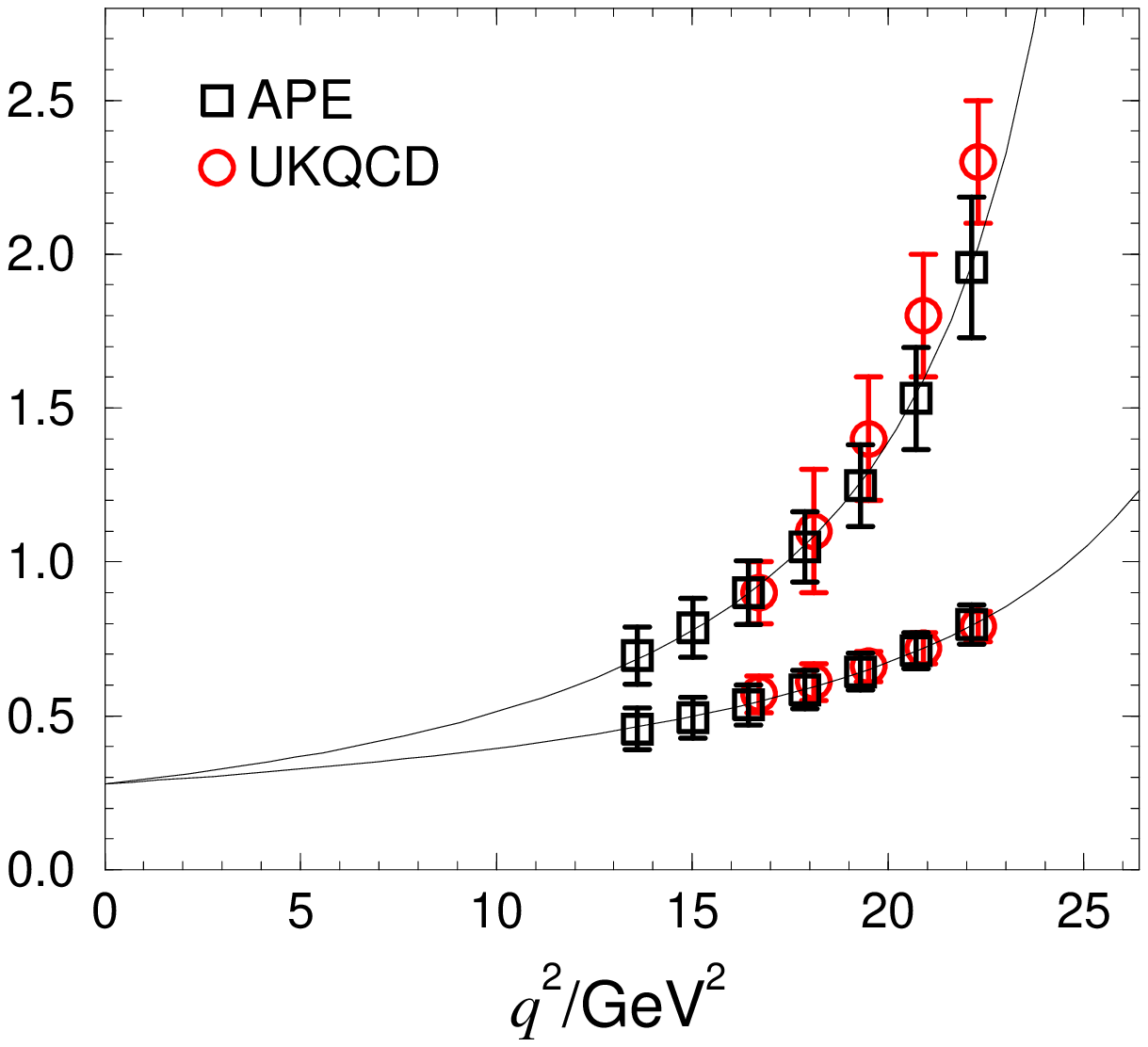}%
\hfill
\includegraphics[width=0.53\hsize]{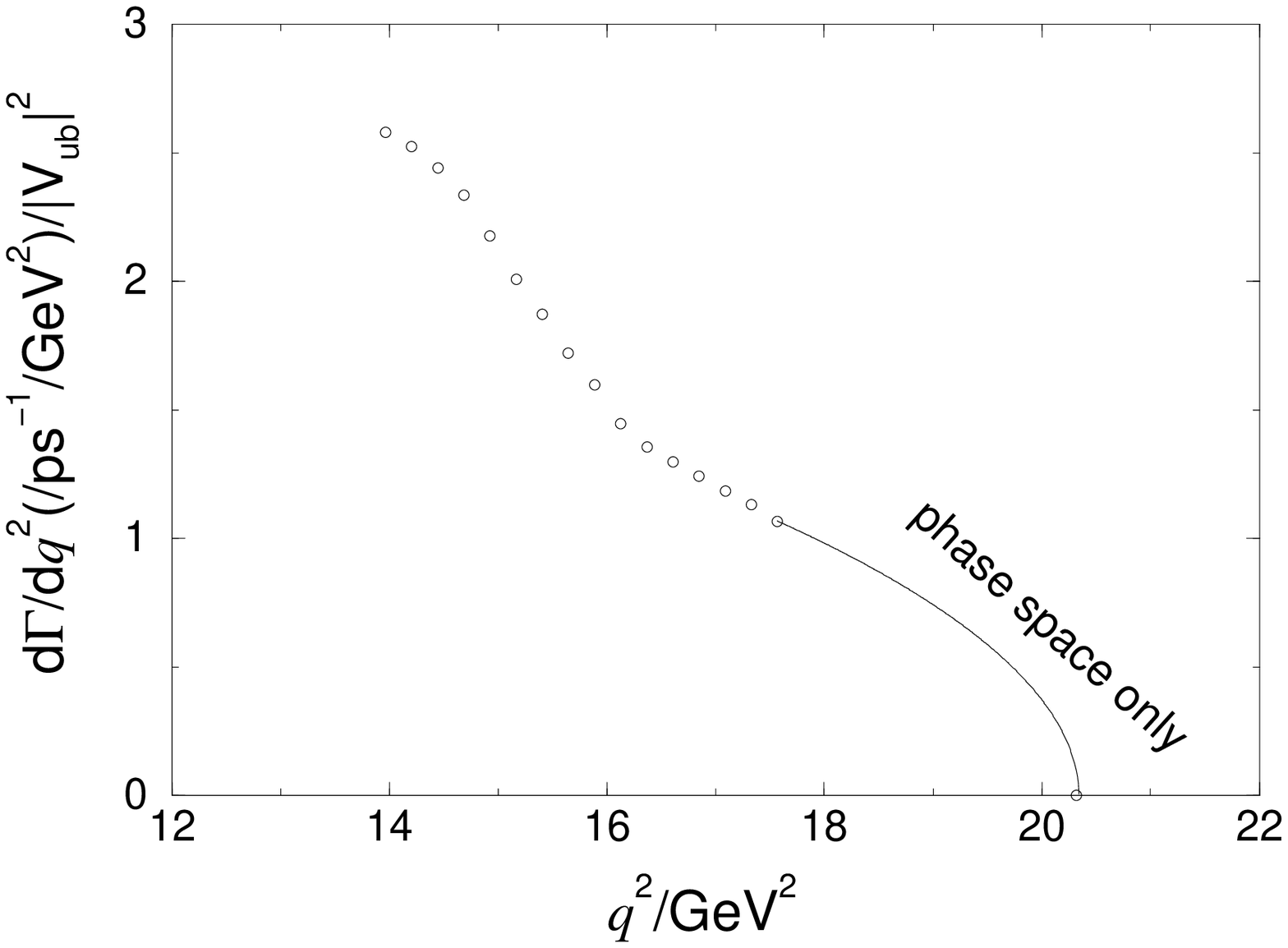}%
}
\caption[]{The left-hand diagram shows results for $f^+$~(upper points)
and $f^0$~(lower points) from the APE~\cite{apesl} and UKQCD~\cite{vil}
collaborations. The curves are fits to the points from APE, showing an
extrapolation to lower values of $q^2$. The right-hand diagram shows the
UKQCD results for the distribution for $b\to\rho$ decays in the range
$14$\,GeV$^2<q^2< 20.3$\,GeV$^2$,
which corresponds to the high $q^2$ bin of CLEO. The region marked
\textit{phase space only} is inaccessible to the lattice calculations
and the curve corresponds to taking the form-factors from the last
lattice point and extrapolating using the phase-space only.
\label{fig:sldata}} \end{figure}

I will only make a few brief comments about semileptonic decays:
\begin{itemize}
\vspace{-0.2in}
\item Lattice simulations of $B\to\pi,\rho$ decays require the momentum
of the final-state light meson to be small in order to avoid
discretization errors. This means that from lattice simulations of
$B\to\pi,\rho$ semileptonic decays we only obtain results at large
values of $q^2$. This is illustrated by the first diagram in
fig.~\ref{fig:sldata}, which contains recent data from the APE~\cite{apesl}
and UKQCD collaborations~\cite{vil}~\footnote{The curves are
APE collaborations fits to their points~\cite{apesl}.}.
\item Lattice computations of semileptonic form-factors have been
performed for many years now. At the lattice 2000 symposium updated
results were presented from the APE, UKQCD, JLQCD and Fermilab
collaborations and ``the results were fairly consistent in the region
where all groups had direct calculations
(19\,GeV$^2<\,q^2\,<\,23\,$GeV$^2$)''~\cite{cb}.  As an example, the first
diagram in fig.\,\ref{fig:sldata} shows the results for the form factors
$f^+$ and $f^0$ for $B\to\pi$ decays.
\item Much effort is being devoted to extrapolating the lattice results 
to smaller values of $q^2$, using as many theoretical constraints as
possible (e.g. HQET scaling relations, unitarity and analyticity,
kinematical constraints, soft pion relations
etc.)~\cite{extrapolations}. The results of such extrapolations for
$f^0$ and $f^+$ are also shown in fig.\,\ref{fig:sldata}.  
\item For a direct application of lattice computations one should
compare the lattice results to experimental distributions at large
values of $q^2$. An example of such a comparison is shown in the second
diagram in fig.\ref{fig:sldata}, where the lattice results from the
UKQCD collaboration for $B\to\rho$ decays~\cite{vil} are 
compared with the high-$q^2$ bin from
CLEO~\cite{cleosl}. Integrating the results in this bin the comparison
takes the form:
\begin{eqnarray}
\frac{\Delta\Gamma(14<q^2/\gev^2<20.3)}
 {\mathrm{ps}^{-1}\gev^{-2}}
 &=&8.3\,|V_{ub}|^2 \hspace{0.64in}\textrm{UKQCD\ preliminary}\\ 
&=&7.1(2.4)\times 10^{-5}\hspace{0.2in} \textrm{CLEO}
\end{eqnarray}
from which one obtains $V_{ub}=2.9(0.5)\times 10^{-3}$.
\end{itemize}

\section{Inclusive Decays and Lifetimes of Beauty Hadrons}
\label{sec:inclusive}

In this section I will briefly mention some physical quantities for
which the first lattice calculations were performed in the last few
years. It will certainly be possible to improve on the precision of
these first-generation computations.

\subsection{Lifetimes of Beauty Hadrons:}\label{subsec:lefetimes}
The fact that the $b$-quark is heavy makes it possible to derive an
operator product expansion for inclusive beauty decays~\cite{bsuv},
which results in an expansion in inverse powers of $m_b$. Specifically
for widths of beauty hadrons:
\begin{equation}
\Gamma(H_b)=\frac{G_F^2\,m_b^5\,|V_{cb}|^2}
{192\pi^3}\left\{c_3\left(1+\frac{\lambda_1+3\lambda_2}{2m_b^2}
\right)+c_5\frac{\lambda_2}{m_b^2}+O\left(\frac{1}{m_b^3}\right)
\right\}\end{equation}
where $c_3$ and $c_5$ are calculable in perturbation theory and
\begin{equation}
-\lambda_1=\frac{1}{2m_{H_b}}\ \langle\,H_b\,|\,\bar h\vec D^2h\,|
\,H_b\,\rangle\ \ \ \textrm{and}\ \ \ 
3\lambda_2=\frac{1}{2m_{H_b}}\ \langle\,H_b\,|\,\frac{g}{2}
\,\bar h\sigma_{ij}G^{ij}h\,|
\,H_b\,\rangle\label{eq:l12def}\end{equation}
defined in the rest frame of $H_b$ ($\lambda_1$ and $\lambda_2$ are the
matrix elements of the kinetic energy and chromomagnetic operators
respectively). This leads to
\begin{eqnarray}
\frac{\tau(B^-)}{\tau(B^0)}&=&1\,+\,O\left(\frac{1}{m_b^3}\right)
\label{eq:taumesons}\\ 
\frac{\tau(\Lambda_b)}{\tau(B^0)}&=&1\,-\,
\frac{\lambda_1(\Lambda_b)-\lambda_1(B^0)}{2m_b^2}
\,+\,3\,c_G\frac{\lambda_2(\Lambda_b)-\lambda_2(B^0)}{m_b^2}+\cdots\\ 
&=&(0.98\pm 0.01)+\,O\left(\frac{1}{m_b^3}\right),
\label{eq:taulambdabth}\end{eqnarray}
to be compared to the experimental values~\footnote{See also
  ref\cite{leptau} in which the result $\tau(B^-)/\tau(B^0)=1.07(2)$
  is quoted.}
\begin{equation}
\frac{\tau(B^-)}{\tau(B^0)}=1.06\pm 0.04\ \ \textrm{and}\ \ 
\frac{\tau(\Lambda_b)}{\tau(B^0)}=0.79\pm 0.05\ .
\label{eq:tauexperiment}\end{equation}
In view of the discrepancy between the theoretical prediction in
(\ref{eq:taulambdabth}) and the experimental results for the ratio
$\tau(\Lambda_b)/\tau(B^0)$ it is important to evaluate the $O(1/m_b^3)$
corrections~\footnote{The significance of the discrepancy is underlined
when one notes that it is really $1-\tau(\Lambda_b)/\tau(B^0)$ which we
calculate.}. These may be significant because it is only at this order
that the light-quark in $B$-meson is involved in the weak decay, and
hence their contribution is enhanced by phase-space factors~\cite{ns}.
To evaluate these $O(1/m_b^3)$ corrections we need to compute the matrix
elements of the following dimension-6 operators~\cite{ns}: 
\begin{eqnarray}
O_1=\bar b\gamma_\mu(1\meno\gamma^5)q\ \bar q\gamma^\mu(1\meno\gamma^5)b,
\ \ &&\hspace{-0.5in}  
O_2=\bar b(1\meno\gamma^5)q\ \bar q(1+\gamma^5)b,\nonumber\\ 
T_1=\bar b\gamma_\mu(1\meno\gamma^5)T^aq\ 
\bar q\gamma^\mu(1\meno\gamma^5)T^ab\ \ &\textrm{and}&\ \ 
T_2=\bar b(1\meno\gamma^5)T^aq\ \bar q(1+\gamma^5)T^ab,
\label{eq:operators}\end{eqnarray}
where $T^a$ represents the colour matrix.

For mesons, the evaluation of these matrix elements is very similar to
that of $B_B$ and is totally straightforward. For the $\Lambda_b$
baryon the calculation is a little less straightforward, but can be
performed using standard techniques. Together with M.~di~Pierro we
found~\cite{massimo}
\begin{equation}
\frac{\tau(B^-)}{\tau(B_d)}=1.03\pm 
0.02\pm 0.03\ ,
\label{eq:taumesonsres}\end{equation}
in good agreement with the experimental result. The first error in
(\ref{eq:taumesonsres}) is the lattice one and the second is our
estimate of the uncertainty due to the fact that the (perturbative)
coefficient function is only known to one-loop. There is a remarkable
factorization of the lattice operators, i.e. the lattice
$B$-parameters are close to 1. For the $\Lambda_b$ baryon we have only
performed an exploratory calculation, in which we did not have
sufficient control of the behaviour with the mass of the
light-quark~\cite{dpms}. The spectator effects were found to be very
significant but, at least for lights quarks corresponding to
$m_\pi\simeq$1\,GeV, the result of $\tau(\Lambda_b)/\tau(B_0)\simeq
0.91$ would imply that spectator effects would not be sufficiently
large to account for the full discrepancy. It is clear however, that
these calculations need to be improved (which will be relatively
simple to do) and that this will make a considerable impact on our
understanding of this important question.

\subsection{\boldmath{$B_s$}-Meson Lifetime Differences:}
\label{subsec:bs}
We now consider the difference of the widths of the $B_s$ mesons. Using
the operator product expansion we have~\cite{bbd}
\begin{eqnarray}
\frac{\Delta\Gamma_{B_s}}{\Gamma_{B_s}}&=&\frac
{G_F^2m_b^2}{12\pi m_{B_s}}\,\left|V_{cb}V_{cs}\right|^2\,\tau_{B_s}\times
\nonumber\\ 
&&\hspace{-0.5in}\Big\{G(z)\langle\overline{B}_s|Q_L(m_b)|B_s\rangle
-G_S(z)\langle\overline{B}_s|Q_S(m_b)|B_s\rangle
+\delta_{1/m}\sqrt{1-4z}\,\Big\}
\end{eqnarray}
where $z=m_c/m_b$, the operators are $Q_L=\bar b\gamma^\mu_Ls\ \bar
b\gamma^\mu_Ls$ and  $Q_S=\bar b(1-\gamma_5)s\ \bar b(1-\gamma_5)s$, the
$\delta_{1/m}$ term represents $O(1/m_b)$ corrections and  $G(z)$ and
$G_S(z)$ have been computed in perturbation theory up to
NLO~\cite{bbgln}. The evaluation of these hadronic matrix elements has
recently began to be performed in lattice simulations by Hashimoto et
al.~\cite{bsjlqcd} and APE collaboration~\cite{bsape}~\footnote{There are also
some even more recent results in the static limit, both in the quenched
approximation and with two flavours of sea quarks~\cite{gr}. The results
quoted in these papers are $5.1\pm 1.9\pm 1.7$\,\% (quenched)
and $4.3\pm 2.0\pm 1.9$\,\% ($n_f=2$), to be compared to those in
eqs.(\ref{eq:bsape}) and (\ref{eq:bsjlqcd}).}. The
APE collaboration quote
\begin{equation}
\frac{\Delta\Gamma_{B_s}}{\Gamma_{B_s}}=(4.7\pm 1.5\pm 1.6)\%\ \ \ 
\textrm{APE}~\cite{bsape},
\label{eq:bsape}\end{equation}
where the second error represents an assumed 30\% uncertainty on the $1/m_b$ corrections.
The calculations were obtained with propagating heavy quarks.
The result from Hashimoto et al. is 
\begin{equation}
\frac{\Delta\Gamma_{B_s}}{\Gamma_{B_s}}=(10.7\pm 2.6\pm 1.4
\pm 1.7)\%\ \ \ \textrm{Hashimoto et al.}~\cite{bsjlqcd},
\label{eq:bsjlqcd}\end{equation}
where the first error is due to the uncertainty in our knowledge of $f_{B_s}$,
the second is the lattice
error in the matrix elements and the third
is the estimate of the error due to the $O(1/m_b)$ corrections.
These results were obtained by simulating NRQCD.

The central values in eqs.~(\ref{eq:bsape}) and (\ref{eq:bsjlqcd}) are
somewhat different (although they are compatible within errors), however
it should be noted that this difference is not due to different lattice
results for the matrix elements. Indeed the values of the matrix
elements determined by the two groups are in good agreement. The
difference is due to the different inputs: Hashimoto et al. use a
lattice result for $f_{B_s}=245\pm 30$\,MeV and the experimental value
of the inclusive semileptonic branching ratio; APE use the lattice value
of $\xi$ and the experimental value of $\tau_{B_s}\,\Delta
m_{B_d}\,m_{B_s}/m_{B_d}$.

\section{The mass of the \boldmath{$b$}-quark}
\label{sec:mb}

The mass of the $b$-quark, $m_b$, is one of the parameters of the
standard model. In this section I will review recent progress in
evaluating  $m_b$ from simulations of effective theories, such as the
HQET or NRQCD~\footnote{For a detailed review of lattice determinations
of quark masses in general, and $m_b$ in particular, see ref.~\cite{vl}.}.
The discussion will be based on the evaluation of the
two-point correlation function
\begin{equation}
C(t)=\sum_{\vec x}\ \langle\,0|\,A_0(\vec x,t)\,A_0(\vec 0,0)\,
|\,0\,\rangle
\label{eq:aa}\end{equation}
in the HQET. $A_\mu$ is the axial current, $A_\mu=\bar
h\,\gamma_\mu\gamma^5\, q$, and  $h$ ($q$) is the heavy-quark
(light-quark) field. The evaluation of $C(t)$ is relatively
straightforward and has been performed for many years now. The progress
in recent years has been in the theoretical understanding of how one can
determine $m_b$ from $C(t)$, and in the perturbative calculations which
have been performed up to three-loop order enabling the results to
be obtained with good precision.

At large times $t$
\begin{equation}
C(t)\simeq Z^2 \exp(-\xi t)\label{eq:casymp}\end{equation}
and from the prefactor $Z$ we obtain the value of the decay constant $f_B$
in the static approximation~\cite{eichtenfb}. It is for this reason that
calculations of $C(t)$ have been performed for some years now.
It was realised later that one can obtain $m_b$ from the measured 
value of $\xi$ (up to $O(\Lambda^2_{\textrm{{\tiny QCD}}}/m_b)$
corrections)~\cite{cgms}. 

The relation between $\xi$ and $m_b$ is a delicate one:
\begin{equation}
M_B=m_b^{\textrm{\tiny{pole}}}+\xi-\delta m\ ,
\label{eq:mbxi}\end{equation}
where $M_B\simeq 5.28$\,GeV is the mass of the $B$-meson,
$m_b^{\textrm{\tiny{pole}}}$ is the pole-mass of the $b$-quark and $\xi$
is determined from lattice computations using eq.~(\ref{eq:casymp}).
$\delta m$ is the residual mass, generated in perturbation theory in the
lattice formulation of the HQET,
\begin{equation}
\delta m = \frac{1}{a}\sum_n X_n\,\alpha_s^n(m_b)\ .
\label{eq:deltam}\end{equation}
In other words even if the action in the HQET is written without a mass
term, higher order terms in perturbation theory generate such a term,
$\delta m$. Each term in the series for $\delta m$ diverges linearly as
$a\to 0$, and these terms (partially) cancel the divergence in
$\xi$. In order for the cancellation to be sufficiently precise the
lattice spacing $a$ cannot be too small.

The perturbation series for $\delta m$ diverges. Indeed it is not
\textit{Borel Summable}, leading to a \textit{renormalon ambiguity},
which is an intrinsic ambiguity of $O(\lqcd)$. The pole mass of a quark
is also not a physical quantity; it also contains a renormalon
ambiguity~\cite{bb,bsuv2}. The renormalons in the pole mass and $\delta
m$ cancel (for a detailed discussion of the cancellation of such
renormalons see ref.~\cite{msrenormalons}). Let $\overline m_b$ be the
$b$-quark mass defined in the $\overline{\textrm{\small{MS}}}$
renormalization scheme defined at the scale $\overline m_b$ itself
(I use $\overline m_b$ as an example of a ``physical'' quark mass). The
pole mass and $\overline m_b$ can be related in perturbation theory:
\begin{eqnarray}
\overline{m}_b&=&\big[1+\sum_n D_n\,\alpha_s^n(m_b)\big]\,m_b^{\textrm{
\tiny{pole}}}\label{eq:mrel}\\ 
&=&\big[1+\sum_n D_n\,\alpha_s^n(m_b)\big]\,
\Big\{M_B-\xi+\frac{1}{a}\sum_n X_n\,\alpha_s^n(m_b)\Big\}\ .
\end{eqnarray}

\subsection{Perturbative Matching}
\label{subsec:mbpert}
All the perturbative coefficients necessary to compute $\overline m_b$ to
$N^3 LO$ have recently been calculated.
\begin{itemize}
\item The coefficients $D_2$ and $D_3$ in the relation (\ref{eq:mrel})
between $\overline m_b$ and the pole mass have been calculated in
refs.~\cite{gbgs} and \cite{mvr} respectively.
\item The relation between the lattice (bare) coupling constant
$\alpha_0$ and the $\overline{\textrm{MS}}$ coupling takes the form
\begin{equation}
\alpha_s(\mu)=\left(1+\sum_n d_n(a\mu) \alpha^n_0
\right)\,\alpha_0.\end{equation}
and the coefficient $d_2$ can be found in refs~\cite{lw,cfpv}.
\item Finally we need to consider the perturbative expansion of the
residual mass:
\begin{equation}
a\delta m = C_1\alpha_0 + C_2\,\alpha_0^2 + C_3\alpha_0^3
+\cdots \label{eq:resmass}\end{equation}
\begin{itemize}
\item $C_1$ is well known, $C_1=2.1173$. 
\item $C_2$ has been determined from the calculation of large
Wilson loops $W(R,T)$, for which
\begin{equation}
W(R,T)\sim\exp(-2\delta m\,(R+T)), 
\end{equation}
where $R$ and $T$ are the spatial and temporal lengths of the
Wilson loop.
\\ $C_2=11.152+n_f(-0.282+0.035c_{\textrm\tiny{SW}}
-0.391c^2_{\textrm\tiny{SW}})$~\cite{ms}, where $c_{\textrm\tiny{SW}}$
is the coefficient of the Sheikholeslami--Wohlert term in the
lattice action~\cite{sw}.
\item The Parma-Milan group has pioneered the numerical computation of
  perturbative coefficients at high orders using the stochastic
  formulation of theory. For the coefficients in
  eq.~(\ref{eq:resmass}) they find~\cite{ados}
\begin{equation}
C_1=2.09(4),\ \ \ 
C_2=10.7(7),\ \ \ \textrm\ \ \  
C_3=86.2(5)\ . \end{equation} 
\item From a numerical Monte-Carlo study on very small and very
fine-grained lattices Lepage at al.~\cite{lmst} find: 
\begin{equation}
C_3=80(2)+5.6(1.8)\ . 
\end{equation}
\end{itemize}
For NRQCD only the first coefficient ($C_1$) is known. 
\end{itemize}
\subsection{Numerical results for \boldmath{$\overline m_b$}}
Before presenting the numerical results I repeat that the evaluation of $\xi$ is
relatively straightforward. The recent progress has been in the evaluation of the
high order perturbative coefficients described in subsection~\ref{subsec:mbpert}. 
The results obtained using the method described in this section have been:
\begin{itemize}
\item\mbox{}\vspace{-25pt}
\begin{equation}
\overline m_b=4.15\pm 0.05\pm 0.20\,\textrm{GeV}\ \ \ \cite{gms},
\end{equation}
obtained in quenched QCD at NLO (i.e. with only the one-loop coefficient
$C_1$). The second error is the estimate of the uncertainty due to
higher order perturbation theory.
\item\mbox{}\vspace{-25pt}
\begin{equation}
\overline m_b=4.30\pm 0.05\pm 0.10\,\textrm{GeV}\ \ \ \cite{ms},\end{equation}
obtained in quenched QCD at N$^2$LO (with coefficients up to $C_2$).
\item\mbox{}\vspace{-25pt}
\begin{equation}
\overline m_b=4.30\pm 0.05\pm 0.05\,\textrm{GeV}\ \ \ \cite{vg},
\label{eq:vg}\end{equation}
obtained in quenched QCD at N$^3$LO (using all the known coefficients
of section~\ref{subsec:mbpert}. 
\item\mbox{}\vspace{-25pt}
\begin{equation}
\overline m_b=4.34\pm 0.03\pm 0.06\,\textrm{GeV}\ \ \ \cite{davies},
\label{eq:davies}\end{equation}
obtained from the static limit of quenched NRQCD at N$^3$LO. 
\item There is also an unquenched result: 
\begin{equation}
\overline m_b=4.26\pm 0.06\pm 0.07\,\textrm{GeV}\ \ \ \cite{ggmr},
\label{eq:ggmr}\end{equation}
obtained in unquenched QCD, with 2 flavours of sea-quarks at 
N$^2$LO. 
\end{itemize}
We take the results in eqs.(\ref{eq:vg}) (or (\ref{eq:davies})) and (\ref{eq:ggmr})
as the current best estimates for $\overline m_b$.

Results obtained using NRQCD are in good agreement with those in the
static theory, but we need to understand the errors due to higher orders
of perturbation theory in that case.

\section{Nonleptonic Decays} \label{sec:nonlept} At this meeting many
interesting results have been presented for exclusive two-body decays of
$B$-mesons. Indeed these decays are one of the principal sources of
information about the unitarity triangle and CP-violation. Apart from
the golden process $B\to J/\Psi K_s$, the determination of the
fundamental parameters from two-body $B$-decays is severely restricted
by our inability to quantify the non-perturbative strong interaction
effects. At present we are some way from understanding how to compute
these effects in lattice simulations, however considerable effort is
being devoted to developing the necessary theoretical framework. Progress
has been made in understanding how to compute $K\to\pi\pi$ decay amplitudes 
\begin{center}
\begin{picture}(300,60)(-150,-30)
\SetWidth{1}
\ArrowLine(-50,0)(0,0)
\ArrowLine(0,0)(50,20)\ArrowLine(0,0)(50,-20)
\GCirc(0,0){5}{0.8}
\Text(-25,10)[b]{$K$}\Text(25,15)[b]{$\pi$}
\Text(25,-15)[t]{$\pi$}
\end{picture}
\end{center}
and I will briefly discuss recent ideas~\cite{mt,ll,lmst2}.
There are two important features which need to be understood:
\begin{enumerate}
\item Lattice calculations are performed in Euclidean space and hence
yield real quantities. One can therefore ask how one can get the full
decay amplitude including the phase due to the final state interactions?
It is true that from each correlation function one obtains a real
quantity, however different correlators give different quantities and
the full amplitude can (at least in principle) be reconstructed. For
example, one can obtain the modulus of the amplitude from one correlator
and the real part from another
which is sufficient to determine the amplitude~\cite{lmst2}.
\item Although momentum is conserved in lattice calculations of
correlation functions, energy is not (correlation functions
are not integrated over time).
\begin{center}
\begin{picture}(185,90)(-50,-50)
\SetWidth{1}
\ArrowLine(-50,0)(0,0)
\Text(-25,7)[b]{$K$}
\Line(0,0)(120,40)\Line(0,0)(120,-40)
\GCirc(0,0){5}{0.8}
\GCirc(120,40){2}{0}\GCirc(120,-40){2}{0}
\Text(0,-10)[t]{$0$}
\Text(128,40)[l]{$t_2$}\Text(128,-40)[l]{$t_1$}
\GOval(70,0)(30,10)(0){0.7}
\Text(30,20)[b]{$\vec p\!=\!0$}
\Text(30,-20)[t]{$\vec p\!=\!0$}
\ArrowLine(105,-35)(106,-35.3333)\Text(105,-41)[t]{$\vec q$}
\ArrowLine(105,35)(106,35.3333)\Text(105,41)[b]{$-\vec q$}
\end{picture}
\end{center}
In this diagram the kaon decays at the origin producing two pions, which
are annihilated at $t_1$ and $t_2$ respectively with momenta $\vec q$
and $-\vec q$. However the pions interact, and can rescatter (as
represented by the grey oval). Whilst the total momentum remains zero,
energy is not conserved and hence there is a contribution in which the
two pions are each at rest (this is the lowest energy state). 
Correlation functions decay exponentially with time as $\exp(-Et)$,
where $E$ is the energy, and hence at large times the dominant
contribution comes from the unphysical matrix element corresponding to a
kaon at rest decaying into two pions, each of which is at rest.
This is a major difficulty for lattice calculations.

In a finite volume the energy levels of the two-pion states are discrete and
L\"uscher and Lellouch have proposed to exploit this fact to determine the
decay amplitude directly~\cite{ll}. One needs to have a state for which
the energy of the two-pion state is equal to $m_K$, the mass of the
kaon, and to be able to isolate the term corresponding to this energy.
In practice, for the near future, this is likely to be the first excited
state.This would require a lattice of about 6\,fm, which is large but not
hugely so~\footnote{The finite volume corrections have also been discussed
in considerable detail in refs.~\cite{ll,lmst2}.}.
\end{enumerate}

The ideas discussed above need to be explored numerically and developed
further. In particular we need to understand how to include the
contributions from inelastic channels (i.e. contributions from
intermediate states other than two-pion ones). The current numerical
studies of  $K\to\pi\pi$ decays are performed on lattices which are
considerably smaller than 6\,fm, and so correspond to unphysical
kinematics. However, one can then use chiral perturbation theory to
obtain the physical amplitudes. For $B$-physics of course chiral
perturbation theory is inapplicable and the neglect of inelastic
channels is not possible, so more theoretical developments are
necessary.

\section{Conclusions}\label{sec:concs}
I conclude with a brief summary of the main points of this talk:
\begin{itemize}
\item Lattice simulations are a central tool in the determination of
information about fundamental physics from experimental data. They are
being applied to a wide range of quantities in $B$-physics.
\item The numerical results presented in
  sections~\ref{sec:fb}\,-\,\ref{sec:mb} are an important part of the
  conclusions.
\item For \textit{standard quantities}, such as $f_B$ and $B_B$, the
  emphasis is on the control of systematic errors, particularly those
  due to quenching. For other quantities (e.g. the applications to
  inclusive decays discussed in section~\ref{sec:inclusive}) the
  calculations are just beginning.
\item For $B\to$ light-meson semileptonic (and radiative) decays, the
matrix elements are computed at large values of $q^2$ and theoretically
motivated extrapolations are performed to lower values of $q^2$. Where
possible, to make optimum use of lattice results, a comparison with the
experimental distributions at large values of $q^2$ should be made.
\item For two-body non-leptonic decays theoretical developments are
needed before lattice simulations can provide information. Progress has
been made for $K\to\pi\pi$ decays and we can look forward to extensions
to $B$-physics.
\end{itemize}

\subsection*{Acknowledgements} I gratefully acknowledge correspondence
with Claude Bernard and Vittorio Lubicz with detailed information
about new results which were presented at Lattice 2000. It is a
pleasure to thank J.Flynn, D.Lin, G.Martinelli and M.Testa for many
instructive discussions. This work was supported by PPARC grant
PPA/G/O/1998/00525.

\end{document}